# Reinforcement Learning-Based Approaches for Enhancing Security and Resilience in Smart Control: A Survey on Attack and Defense Mechanisms

Zheyu Zhang, Virginia Tech

February 23, 2024

[1]

**Abstract**

Reinforcement Learning (RL), one of the core paradigms in machine learning, learns to make decisions based on real-world experiences. This approach has significantly advanced AI applications across various domains, notably in smart grid optimization and smart home automation. However, the proliferation of RL in these critical sectors has also exposed them to sophisticated adversarial attacks that target the underlying neural network policies, compromising system integrity. Given the pivotal role of RL in enhancing the efficiency and sustainability of smart grids and the personalized convenience in smart homes, ensuring the security of these systems is paramount. This paper aims to bolster the resilience of RL frameworks within these specific contexts, addressing the unique challenges posed by the intricate and potentially adversarial environments of smart grids and smart homes. We provide a thorough review of the latest adversarial RL threats and outline effective defense strategies tailored to safeguard these applications. Our comparative analysis sheds light on the nuances of adversarial tactics against RL-driven smart systems and evaluates the defense mechanisms, focusing on their innovative contributions, limitations, and the compromises they entail. By concentrating on the smart grid and smart home scenarios, this survey equips ML developers and researchers with the insights needed to secure RL applications against emerging threats, ensuring their reliability and safety in our increasingly connected world.

## 1 Introduction

ML technologies render the computer to understand the information like a human that can figure out what they saw and what they hear. It has achieved great success in computer vision, intelligence systems, natural language processing, etc[24, 44, 16]. Meanwhile, Reinforcement learning is a unique ML technology that lets computers take actions like human beings. The reason behind the action is to learn from real-world experience or self-play results.

The field of reinforcement learning (RL) research is continually evolving, extending its reach across various domains. Enabled by advancements in deep neural networks [1], RL has been successfully applied to complex robotics control tasks [13]. Zhu et al. [51] have employed RL models for targetdriven visual navigation, illustrating the versatility of RL applications. Drawing inspiration from the human brain, a novel approach termed neuro-evolution has emerged [34], integrating neuro-evolutionary strategies with RL to forge advanced deep RL systems that more closely mimic human cognitive processes. In the realm of autonomous driving, deep reinforcement learning has been instrumental in developing long-term self-driving strategies, effectively learning control policies from high-dimensional inputs [27, 23]. Furthermore, RL is emerging as a promising technology for Connected Automated Vehicle (CAV) systems, offering significant potential to revolutionize transportation systems in the future [8]. A notable achievement in the field was the Google DeepMind team's victory over world champion Lee Sedol in Go, achieved with AlphaGo [31], trained using an RL paradigm. This was followed by AlphaZero [32], which achieved superhuman performance in a self-play training style. In 2018, OpenAI Five demonstrated the prowess of RL by using five Long Short-Term Memory networks and a Proximal Policy Optimization method [26] to defeat a professional Dota team. The surge in Multi-Agent Reinforcement Learning (MARL) research [50] is testament to the growing interest in this area, with deep neural networks enabling MARL to tackle complex problems that surpass traditional challenges, even in real-world scenarios.

---

[1] Contact: zheyuzhang21@vt.edu, Virginia Tech, 900 N Glebe Rd, Arlington, VA, USA, 22203.

In order to strengthen the security application of deep reinforcement learning in the variant architecture of machine learning[20, 39, 28, 48], we should find the loopholes of the deep reinforcement learning algorithm as soon as possible and prevent malicious users from using these loopholes to make illegal profits Different from the traditional one-step prediction task of machine learning.[15] deep reinforcement learning system uses multi-step decision-making to complete specific tasks[15], and there is a high correlation between continuous decisions Generally speaking, the attack of deep reinforcement learning system can launch malicious attacks on five main links of reinforcement learning algorithm, including environment, observation, reward, action and strategy[37],[21], Therefore, before the deep reinforcement learning system is really applied to the actual industry, it is very important to explore the vulnerability of the deep reinforcement learning system and improve its defense ability and robustness In order to improve the robustness of deep learning model[37], many DRL defense methods have been proposed, mainly including three directions: countermeasure training, robust learning and countermeasure detection At present, there is still much room for development in the research of attack and defense in the field of deep reinforcement learning[12]. In view of the problems that deep reinforcement learning is vulnerable to counter sample attacks, the robustness optimization of deep reinforcement learning model and counter-defense methods have also become the focus of attention, which still needs to be explored At the same time, due to the application of deep reinforcement learning in security related fields, its strategic loopholes have also become a major security hidden danger[40] In order to better explore the attack and defense research status and future development direction of deep reinforcement learning system, this paper summarizes the deep reinforcement learning algorithm, attack and defense methods, and security analysis as comprehensively as possible[20].

The main proposes of this survey are:

• attempts to convey a comprehensive survey on state-of-the-art adversarial RL attacks and summarize the typical defense methods to mitigate or against such potential threats for RL systems. We horizontally compare the variant characteristics of adversarial attack mechanisms to typical RL systems. • review the related defense technologies with their technical innovations, drawbacks, and trade-offs. We believe this survey will provide ML developers and researchers with a security angle to facilitate the RL system's reliability and safety for the upcoming era with ubiquitous RL.

Survey structure: in the background, we describe the terminologies related to adversarial attacks to RL, and introduce the important reinforcement learning algorithms. The attack part of RL reviews the relevant research on antagonistic attacks in the RL. The defense technology of antagonistic attack under the background of reinforcement learning, the defense technology of antagonistic attack. We have a general review in the conclusion and discussion.

## 2 Background and Related Work

### 2.1 Reinforcement Learning

A typical RL system descript a task as a Markov decision process (MDP)[17]. It is a mathematical paradigm for modeling the process that takes action in an environment and learning the experience by receiving a reward from each state in the Markov state. The goal of an RL system is to maximize the expectation of accumulated rewards[17, 36],.

Firstly, define the MDP with time step $t \in \{0,1,...,T\}$, the agent is in the state of environment $s_t$ and has an action $a_t$. Then the agent decides to go to a new state $s_{t+1}$ and gets a corresponding reward $r(s_t,a_t)$. The target of the RL system is to find a policy $\pi(a_t|s_t)$ for tuning the action to maximize the utility function $J_\theta(\pi)$.

$$J_\theta(\pi) = E_{s_0,a_0,...}[\sum_{t=0}^{\infty} \gamma^t r(s_t, a_t)]$$

In utility function, $\gamma$ is a discount factor that $\gamma \in [0,1]$; $a_t$ base on $\pi(s_t,a_t)$ is the action drawn from the given policy $\pi(s_t,a_t)$ and $s_{t+1}$ base on $P(s_t+1|s_t,a_t)$ is correspnding to the dynamics environment[19].

$$V^\pi = E_{s_{t+1}, a_t, \ldots}[\sum_{t=0}^{\infty} \gamma^t r(s_{t+i}, a_{t+i})]$$

V-function is the The state value function expected return by policy p from state[19].

$$Q^\pi(s_t, a_t) = E_{s_{t+1}, a_{t+1}, \ldots}[\sum_{t=0}^{\infty} \gamma^t r(s_{t+1}, a_{t+1})]$$

Q-function is the expected return by policy p after taking action at at state[19].

## 2.2 Adversarial Examples Attack and Defense

Adversarial examples can cause AI systems to be deceived[12]. The general form of countering samples is to add small perturbations to information carriers such as images and voices, which are still difficult to detect for the human visual system. Implicit AE refers to a upgraded version of the validation Information Vector that confuses or deceives machine learning technology by adding invisible human disturbances to the global data at the single-pixel-level[37]. Furthermore, the DAE is a upgraded version of the Clean-Map that alters local information by adding physical barriers to confuse the Enhanced Learning-based routing algorithm[15].

- Adversarial Example Attacks: Recent scholarly research has unveiled that deep learning models are susceptible to adversarial example attacks, posing significant challenges to their security, robustness, and adaptability. An adversarial example is a strategically modified input that, while seemingly similar to the original, leads the model to misclassify it with high confidence. The primary objective is not merely to generate and deploy adversarial examples to undermine the model but to enhance the model's resilience by training it to recognize both original and adversarial inputs. This approach aims to formulate a minimax problem, bolstering the model's robustness and security. Initially focused on computer vision, this line of research has expanded into Natural Language Processing.

  The vulnerability of Reinforcement Learning systems, especially those integrated into networked infrastructures like smart grids and smart homes, spans extensively across the network domain. This encompasses risks associated with network protocols, the diverse frequencies used by RL systems, and the threats from unreliable service providers. Weaknesses in communication protocols[18] can be exploited to intercept or alter data flow between RL agents and their environments, compromising the decision-making process. The dependence on wireless communications exposes RL systems to spectrum-based attacks[30, 43, 41, 42], such as jamming or spoofing[29], potentially crippling their functionality. Moreover, the involvement of untrusted or compromised service providers magnifies the threat landscape, facilitating malicious activities like data tampering or unauthorized data access[45, 7, 6]. It is imperative to address these complex network vulnerabilities to safeguard the integrity, confidentiality, and availability of RL applications in critical settings.

  This article synthesizes the existing literature and our insights on the application of adversarial machine learning in recommendation systems, a domain where the balance between data providers and consumers is critical. Recommendation systems, particularly those employing latent factor models like matrix factorization and deep collaborative filtering, have gained prominence for their superior performance and precision. Despite their success, recent findings indicate these systems' vulnerability to adversarial attacks, compelling the recommendation models to err. Consequently, integrating adversarial training with machine learning techniques to counter adversarial examples has garnered significant interest in recent years.

- AE defense: After introducing the relevant attack categories, we roughly introduced the protection methods of the defense attacks. The methods to prevent the model from being attacked were mainly through the data collection stage, the data training stage, and the reasoning stage. The techniques involved included anonymousness, differential privacy, multiple security calculations, and the same state encrypted techniques.

The implementation of defense mechanisms against adversarial attacks, particularly in complex systems like smart grids and smart homes, can be prohibitively expensive due to the computational and resource-intensive nature of these solutions. However, leveraging heterogeneous computing architectures offers a promising avenue to mitigate these costs[49]. By utilizing a diverse array of computational resources, including CPUs, GPUs, and FPGAs, it's possible to accelerate cryptographic algorithms and other security protocols, thereby enhancing the efficiency and feasibility of defense strategies[9, 46]. This approach allows for the distribution of computational tasks to the most suitable processing units, optimizing performance and energy consumption[47, **?**]. Heterogeneous computing can thus play a pivotal role in enabling robust security measures without imposing unsustainable costs, ensuring the practical implementation of advanced defense mechanisms in protecting against adversarial threats.

In addition, more and more attention has been paid to the latest technology such as machine learning and federal learning. Many things in the world have two sides. The advantage of one side could become the disadvantage of the other side, and so did the machine learning and deep learning models. When the model relied on advanced calculations to feed the data constantly, although it could dig out the accurate behavior preferences of the users, it would undoubtedly increase the risk of leaking the privacy data of the users. Therefore, how to weigh the effectiveness of big data and privacy security is a question worth our in-depth study. I hope the conclusion can be helpful to you[24].

## 2.3 Threat model

Almost all of the loopholes lay in that the RL system could not understand the contents of its input from the meaning. For example, the AEs are inputs with similar meanings, but they would cause different predictions. Any RL system that was composed of the RL could be attacked in a way similar to the DNN. We define the attack types in this article and list the basic threat model for further analysis[38].

- white-box attack: The attacker breaks the confidentiality of the architecture and weights of the RL system. • Black-box attack: The attacker has no attack surface to access the architecture and weights of the RL agents.

- Training phase attack: attackers could choose what kind of input the system would give during the training to influence the training by changing the value of the input.

- Inferring phase attack: attackers could choose which input was provided to a well-trained system to influence the system's behavior by changing the value of the sensor or directly providing the input.

# 3 Summary of Attack Methods in Reinforcement Learning

In this section, we explore significant research aimed at enhancing the adversarial robustness within the realm of reinforcement learning (RL). Our review primarily focuses on strategies to combat specific adversarial scenarios. By introducing perturbations to the informational medium, adversarial examples are generated, enabling a counteractive response to bolster the RL system's defenses. We delve into works addressing defensive measures against RL systems under white-box attack scenarios and extend our discussion to black-box attacks, evaluating the practicality and contributions of adversarial research in these domains. Furthermore, we encapsulate the characteristics of various defensive attack methodologies examined in this segment.

**Fast Gradient Sign Method (FGSM)**[12]: This method introduces perturbations in the direction of the loss function gradient, leading to misclassification by the target model. Specifically designed for efficiency, the FGSM modifies the input to resemble the loss function, exploiting the linear characteristics of artificial neural networks (ANNs) to enhance vulnerability to such disturbances. The adversarial example generation using FGSM is formalized as $\eta = \epsilon \cdot \text{sign}(\nabla_x J(\theta,x,y))$, where $\theta$ represents model parameters, $x$ the input, $y$ the output, and $J$ the cost function.

| Method | Threat model and Target algorithm | Application scenario | Attack effect |
| --- | --- | --- | --- |
| Fast gradient sign method | White Box/None | Atari Game | Taking wrong action |
| Start point-based adversarial attack on Q-learning | White Box/Q-Learning | Path Planning | No normal path planning |
| Carlini and Wagner | White Box/None | Path Planning | No normal path planning |
| Common Dominant Adversarial Examples Generation Method | White Box/A3C | Path Planning | Unable to reach destination/Time increased |
| DeepFool | White Box/None | Atari Game | Taking wrong action |
| Basic Iteration Method | Black Box/DQN | Atari Game | Taking wrong action |
| Jacobi-based Significant Map Attack | White Box/None | Atari Game | Taking wrong action |
| Enchanting Attack | Black Box/None | Atari Game | Taking wrong action |
| Adversarial attack on VIN | Black Box/VIN | Path Planning | No normal path planning |

Figure 1: Summary table of defensive attack methods discussed, highlighting application areas and their effectiveness in RL systems under attack.

**Start Point-based Adversarial Attack**[35]: This approach, known as defensive Q-learning, employs a probabilistic model to predict adversarial examples based on influential factors and their associated weights. Factors such as the energy point gravity, key point gravity, path gravity, and angular considerations are integrated through PCA-weighted analysis to construct a predictive model.

**Carlini and Wagner Attack (C&W)**[5]: The C&W attack optimizes an objective function to ensure misclassification, evaluating different loss functions to identify the most effective approach. The chosen function evaluates the proximity of the adversarial example $X'$ to being classified as a targeted class by analyzing the logits or probability predictions of the model.

**Common Dominant Adversarial Examples Generation Method (CDG)**[2]: The CDG focuses on generating adversarial examples particularly effective in disrupting A3C pathfinding algorithms. By identifying and exploiting the most vulnerable aspects of the model, this method introduces disturbances that significantly impair the RL agent's performance.

**DeepFool**[22]: This technique offers a straightforward and precise way to gauge the model's susceptibility to perturbations. DeepFool iteratively projects the input onto the nearest decision boundary, thus crafting minimally altered adversarial examples that lead to misclassification.

**Basic Iterative Method (BIM)**[10]: An extension of FGSM, BIM applies small, iterative adjustments to craft adversarial examples within the vicinity of the original input, maintaining the perturbation's magnitude under a predefined threshold $\epsilon$.

**Jacobi-based Saliency Map Attack (JSMA)**[11]: JSMA highlights critical features contributing to misclassification through a saliency map, iteratively modifying the input to enhance the likelihood of incorrect predictions.

# 4 Summary of Defense Methods in Reinforcement Learning

This section presents innovative defense strategies from three perspectives: input adjustment, target function optimization, and network structure enhancement.

| Defense Method: Tunning Inputs Defense | Effective to Attack | Defense Method: Optimizing Rewards Function | Effective to Attack | Defense Method: Enhance Network Architecture | Effective to Attack |
|---|---|---|---|---|---|
| Adversarial training | FGSM | Adding stability term | FGSM | Defensive distillation | FGSM, BIM |
| Ensemble adversarial training | FGSM | Adding regularization term | DeepFool | High-level representation Guided Denoiser | FGSM, BIM |
| Cascade adversarial training | BIM | Dynamic quantized activation function | DeepFool, FGSM, PGM | Add detector subnetwork | FGSM, BIM DeepFool |
| Principled adversarial training | PGM | Stochastic activation pruning | FGSM | Generative models | FGSM, C&W |
| Gradient band-based adversarial training | CDG | | | Characterizing adversarial subspaces | FGSM, C&W |
| Data randomization | DeepFool, C&W | | | Defensive distillation | FGSM, CDG, DeepFool, JSMA, |
| Input transformations | DeepFool, C&W | | | MAND | FGSM, BIM JSMA,C&W, DeepFool |
| Input gradient regularization | JSMA, BIM | | | | |

Figure 2: Summary table of defensive methods discussed, detailing adjustments to inputs, reward function optimization, network architecture enhancements, and their effectiveness against specific attacks.

# 5 Detailed Defense Mechanisms in Reinforcement Learning

## 5.1 Adjusting Inputs

Adjusting inputs is a critical defense strategy aimed at enhancing the robustness of machine learning models against adversarial attacks. This approach primarily focuses on the preprocessing of inputs to make models less susceptible to minor perturbations that could lead to misclassification or erroneous predictions. Techniques such as *adversarial training* involve exposing the model to adversarial examples during the training phase, thereby 'inoculating' the model against future attacks by expanding its training dataset to include these manipulated inputs. This method not only improves the model's accuracy in the face of adversarial inputs but also contributes to a more generalized and resilient model performance[39, 25, 14].

Another significant technique within this domain is *input conversion*, which transforms the inputs in a way that reduces the effectiveness of adversarial perturbations. Common methods include adding noise, compressing images, or applying other transformations that are likely to negate the changes introduced by an adversary. These transformations can significantly degrade the potency of adversarial examples without substantially impacting the model's ability to make accurate predictions on legitimate inputs.

## 5.2 Optimizing Reward Functions

In reinforcement learning (RL), optimizing reward functions as a defense mechanism involves modifying the reward structure of the RL system to diminish the impact of adversarial actions. The reward function, which guides the learning process by assigning value to the outcomes of different actions, becomes a critical component in ensuring the system's resilience to adversarial manipulations[33, 3, 4].

By carefully designing reward functions to be less sensitive to adversarial perturbations or by incorporating mechanisms that detect and penalize actions leading to adversarial states, systems can maintain their intended performance even in the presence of adversarial interference. This might involve complex reward shaping techniques, where additional rewards or penalties are introduced to guide the agent away from states that could be exploitable by adversaries.

Moreover, robust reward functions can be designed to account for uncertainty and adversarial actions, ensuring that the RL agent remains focused on long-term goals rather than being misled by short-term adversarial gains. This approach not only enhances the system's resistance to attacks but also contributes to the development of more sophisticated and adaptive RL agents capable of operating in dynamic and potentially adversarial environments.

### 5.3 Enhancing Network Architecture

Enhancing the network architecture involves structural modifications and training methodologies that increase the neural network's resilience to adversarial examples. *Adversarial training*, one of the most effective strategies in this category, iteratively exposes the network to adversarial examples alongside genuine inputs during the training phase. This exposure helps the network learn to identify and correctly classify adversarial inputs, thereby reducing their impact[33, 25].

Additionally, techniques like *dynamic activation function quantization* and *randomly activated pruning* introduce variability and redundancy within the network's architecture, making it more difficult for adversarial inputs to exploit consistent vulnerabilities. Dynamic quantization adjusts the precision of the activation functions in response to detected adversarial attempts, while randomly activated pruning temporarily disables random subsets of neurons, forcing the network to rely on multiple pathways for decision-making, thereby enhancing its robustness.

These architectural enhancements are crucial for developing ML models capable of withstanding sophisticated adversarial attacks without compromising their performance on legitimate tasks. By integrating these defense mechanisms, researchers and practitioners can significantly improve the security and reliability of RL systems in various applications.

## 6 Conclusion

RL, a fundamental ML paradigm, has recently revolutionized various AI applications. Despite its advancements, the susceptibility of RL systems to adversarial attacks necessitates enhanced reliability, particularly in critical environments like military applications. This paper provides a comprehensive analysis of advanced defensive strategies against RL attacks, highlighting their characteristics, innovations, and limitations. Our examination aims to offer a security-oriented perspective to ML practitioners, contributing to the development of more secure and reliable RL systems in the foreseeable future of ubiquitous RL applications.